\documentclass[aps, prd, twocolumn, superscriptaddress, nofootinbib, showpacs]{revtex4}
\usepackage[dvips]{graphicx}

\def\lsim{\lower0.6ex\vbox{\hbox{$ \buildrel{\textstyle <}\over{\sim}\ $}}}
\def\gsim{\lower0.6ex\vbox{\hbox{$ \buildrel{\textstyle >}\over{\sim}\ $}}}

\def\beq{\begin{equation}}
\def\eeq{\end{equation}}

\def\bfig{\begin{figure}[t] \begin{center}}
\def\efig{\end{center} \end{figure}}

\def\Om{$\Omega_{\rm M}\ $}

\def\Or{$\Omega_{\rm R}\ $}

\def\Neff{$N_{\nu ,eff}\ $}
\def\lcdm{$\Lambda$CDM }

\def\etal{{\it et al.},~}

\def\ie{{\it i.e.},~}

\def\4he{$^4$He}

\def\7li{$^7$Li}

\begin{document}

\title{Constraints on the cosmological relativistic energy density}

\author{Andrew R. Zentner}
\email{zentner@pacific.mps.ohio-state.edu}
\affiliation{Department of Physics, The Ohio State University,
Columbus, OH 43210, USA}
\author{Terry P. Walker}
\email{twalker@pacific.mps.ohio-state.edu}
\affiliation{Department of Physics, The Ohio State University, 
Columbus, OH 43210, USA}
\affiliation{Department of Astronomy, The Ohio State University,
Columbus, OH 43210, USA}

\date{December 17, 2001}

\begin{abstract}

We discuss bounds on the cosmological relativistic energy density as 
a function of redshift, reviewing the big bang nucleosynthesis  and
cosmic microwave background bounds, updating bounds from large scale structure,
and introducing a new bound from the magnitude-redshift relation for Type Ia
supernovae. We conclude that the standard and well-motivated assumption that 
relativistic energy is negligible during recent epochs is not necessitated 
by extant data.   We then demonstrate the utility of these bounds by constraining 
the mass and lifetime of a hypothetical massive big bang relic particle.  

\end{abstract}


\maketitle
\section{\label{sec:Intro}Introduction}

Cosmological constraints on the relativistic energy density 
(RED) of the Universe, particularly the bound derived from the successful 
prediction of the primordial abundances of D, \4he, and \7li, provide perhaps 
the strongest connections between cosmology and particle physics.  
In this paper we examine constraints on the RED (To be specific, we consider 
constraints on the energy density of any particle that obeys an equation of 
state $P=\rho/3$.) at various cosmological epochs and discuss how these bounds 
can be used to limit the mass and lifetime of an unstable big bang relic particle.  
We review the classic big bang nucleosynthesis (BBN) bound and the bound from the 
cosmic microwave background (CMB) anisotropy power spectrum.  We present a new 
bound derived from the magnitude-redshift relation of Type Ia supernovae (SNIa), 
and we update the bound from large scale structure (LSS).  Although we will attempt 
to derive bounds that are independent of additional cosmological parameters, where 
necessary we will borrow constraints from other techniques in order to further 
limit the relativistic energy density of the Universe.  For example, when needed 
we adopt the bound $0.1 \le \Omega_{\rm M} \le 0.5$ ($\Omega_{\rm M}$ is the matter 
density relative to the critical density today, $\rho_{\rm crit} \equiv 3H_0^2/8\pi G$), 
consistent with recent measurements of the contemporary matter density from lensing 
\cite{LENS}, the rich cluster baryon fraction \cite{F_B}, mass-to-light ratio 
estimates \cite{BAHCALL}, and the power spectrum of the Ly$\alpha$ forest \cite{LYA}.  
We also adopt a hard limit on the Hubble parameter ($h \equiv H_0/100$ kms$^{-1}$Mpc$^{-1}$) 
of $0.56 \le h \le 0.88$ in accordance with the $2\sigma$ range of the HST Key 
Project \cite{HST}. 

\section{\label{sec:CRE}Constraints on Relativistic Energy}

\subsection{Nucleosynthesis Constraint}

It has long been known that an increase in the RED of the Universe during BBN
leads to an overproduction of \4he \cite{SSG77} (BBN reviews 
include \cite{BBNREV}).  Prior to freeze-out of the weak interactions, the 
neutron-to-proton ratio $(n/p)$ tracks its equilibrium value. As the 
Universe expands and cools, the weak interactions freeze-out and $(n/p)$ 
freezes in.  Freeze-out occurs when the weak interaction rates 
become comparable to the expansion rate $H$, given by 
$H^2 = 8\pi G\rho/3$, where $\rho$ is the mean cosmological energy density.  
Once deuterium can be synthesized nearly all available neutrons 
are incorporated into \4he.  Increasing the energy density of the Universe 
during BBN increases the expansion rate, causing the weak interactions to 
freeze-out at a higher temperature and therefore, at a higher value of $(n/p)$.  
An increased expansion rate also allows less time for free neutron decay.  
Both effects make more neutrons available for \4he synthesis.  

In addition, BBN production of \4he slowly increases with the baryon density or equivalently, 
the baryon-to-photon ratio, $\eta \equiv n_B/n_{\gamma}$.  Since BBN 
production of deuterium is a rapidly decreasing function of $\eta$, the baryon 
density can be deduced from the D/H ratio observed in QSO absorption line systems.  
Furthermore, because \4he has been produced in stars since BBN, 
the observed \4he abundance places an upper limit on its primordial abundance.  
Armed with this upper bound to primordial \4he and an independent measure 
of $\eta$, we can place a bound on the RED during the BBN epoch ($z\sim10^9$).  
\begin{figure}
\resizebox{!}{7cm}
{\includegraphics{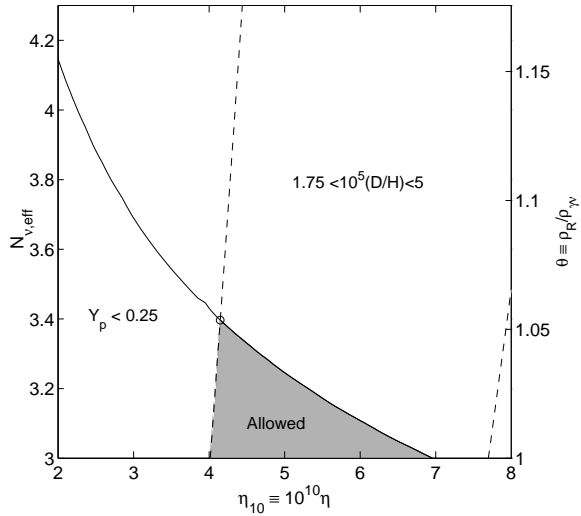}}
\caption{\label{fig1} The BBN bound on the RED.  The region bounded by the dashed curves 
delimits the parameter space for which the theoretically predicted and observed values of 
D/H are consistent.  Likewise, the data on $Y_P$ are consistent with predictions
in the region below the solid curve.  The intersection, marked by the circle, shows
the maximum value of \Neff that can be made consistent with these limits on $Y_P$ and D/H.}
\end{figure}

\begin{figure}
\resizebox{!}{7cm}
{\includegraphics{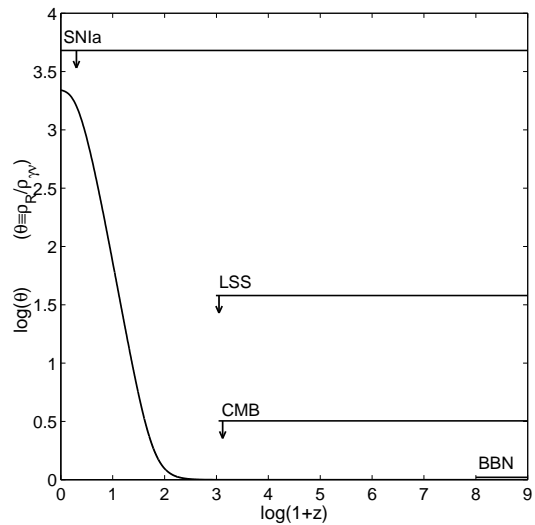}}
\caption{\label{fig2} Constraints on the RED.  The horizontal lines denote the maximum
value of $\theta(z)$ allowed by BBN, CMB, LSS and SNIa.  Notice that the BBN constraint 
in the lower right-hand corner is hardly noticeable (\ie very strict).  By extending the 
limits to higher redshift, we have made the assumption that after BBN relativistic 
energy is only injected and never removed.  The curved line shows $\theta(z)$ 
in a scenario where a massive big bang relic with $M_{keV}Y = 1.2 \times 10^{-3}$ decays 
with lifetime $\tau_{yr} = 2 \times 10^9$.  There are additional CMB and 
LSS constraints on relativistic energy injected after $z \sim 1000$; however, 
they do not rule out the decay shown above (see sections \ref{sec:WIMP} 
and \ref{Con}).}
\end{figure}

We illustrate the current status of the BBN bound to the RED as follows (see
the recent analyses of \cite{BBNLIMIT} for details).  
We adopt a liberal range for the primordial D/H ratio that accommodates 
recent observations of deuterium in several QSO absorption line systems 
\cite{DOBS}:  $1.75 \le 10^{5}\textrm{(D/H)} \le 5.0$.  We then bound the 
RED by taking a conservative bound on the \4he mass fraction, $Y_P \le 0.25$.
This bound is consistent with both the low \cite{OSS97} and high \cite{IT98} 
zero-metallicity extrapolations of low-metallicity HII region data (but see \cite{ADDY}). 
We follow convention by expressing limits on relativistic energy in terms of the energy 
carried by an effective number of light neutrino species, $N_{\nu ,eff}$.  
The RED at some redshift $z$, after e$^+$e$^-$ annihilation, is related to 
$N_{\nu,eff}$ by  
\beq
\theta(z) \equiv \frac{\rho_{\rm R}(z)}{\rho_{\rm{\gamma}{\nu}}(z)} 
= [1 + 0.135(N_{\nu,eff} - 3)],
\eeq
where we scale the RED by 
$\rho_{\rm{\gamma}{\nu}}(z)$ which is the standard model RED carried by 
photons and three massless neutrino species 
(with a current CMB temperature of $2.725$ K \cite{TO1999}, 
$\Omega_{\rm \gamma\nu} \equiv \rho_{\rm \gamma\nu}(z=0)/\rho_{\rm crit} 
= 4.18h^{-2} \times 10^{-5}$).  In Figure \ref{fig1}, 
we show regions of the $\eta$-\Neff plane allowed by the observed \4he and D abundances.  
The data require that $N_{\nu ,eff} \lsim 3.4$ or, in terms of the RED during 
nucleosynthesis $\theta_{\rm BBN} \equiv \theta(z=10^9)$,
\beq
\label{eq:eq5}
\theta_{\rm BBN} \lsim 1.05.
\eeq
In Figure \ref{fig2}, we show the amount of extra relativistic energy permitted during 
BBN.

There are two important caveats regarding the BBN RED bound.  First, although
(\ref{eq:eq5}) is a rather stringent bound on the RED, corresponding to an 
increase of $\sim 5\%$ over the standard $\rho_{\rm \gamma\nu}$, it only applies 
during the epoch of nucleosynthesis.  Relativistic energy injected after BBN 
is not subject to this bound (in Section \ref{sec:WIMP}, we discuss a toy model 
in which relativistic energy is injected into the Universe after the light nuclide 
abundances have been fixed).  Second, the BBN RED bound 
is flavor dependent in the sense that extra relativistic energy in the 
form of degenerate electron neutrinos changes the rates of the weak 
interactions that inter-convert protons and neutrons thereby changing 
theoretical BBN yields.  Introducing an appropriate amount of electron
neutrino degeneracy can always compensate for excess relativistic energy
present during nucleosynthesis \cite{DEGRED, KSSW}, nullifying the BBN
RED constraint.    

\subsection{\label{sub:CMB}CMB Constraint}

The CMB anisotropy power spectrum is also sensitive to the RED 
of the Universe at recombination, primarily through the early Integrated 
Sachs-Wolfe (ISW) effect.  The erosion of gravitational potentials due to incomplete 
matter domination at recombination leads to a boost in power, particularly near the first 
acoustic peak (for a review see \cite{CMBREV}).  CMB constraints on relativistic energy 
complement those from BBN in two ways.  First, the CMB constraints are 
flavor independent; the power spectrum measures relativistic energy regardless
of its form.  Second, the CMB constrains the RED at later epochs 
so that the CMB can be used to study the injection of relativistic energy
after the epoch of nucleosynthesis, for example, by massive neutrino decays.  

Wang, Tegmark \& Zaldarriaga \cite{WTZ01} have compiled a combined CMB data 
set including the recent results from the BOOMERANG \cite{BOOM}, 
MAXIMA \cite{MAX} and DASI \cite{DASI} experiments.  Hannestad \cite{HANN01} has 
performed a likelihood analysis on this combined data set and found, with weak priors
on the Hubble and tilt parameters and the baryon density, $N_{\nu ,eff} \le 19$ with 
95\% confidence.  In terms of the RED present at recombination, 
$\theta_{\rm CMB} \equiv \theta(z \approx 1100)$, this bound is 

\beq
\label{eq:eq6}
\theta_{\rm CMB} \lsim 3.2
\eeq 
as depicted in Figure 2.

A cautionary note is in order.  Although the priors chosen by 
Hannestad are quite conservative, it must be borne in mind that the bound in 
(\ref{eq:eq6}) does depend upon these priors.  For example, by adopting tighter, 
yet reasonable, prior constraints on $h$ and $\eta$, Hannestad shrinks the above 
limit to $N_{\nu,eff} \le 17.5$ at 95\% confidence \cite{HANN01} (see also 
\cite{HMMMP01}).  Kneller \etal \cite{KSSW} have explored the prior dependence 
of these bounds in detail, showing that the bound on $N_{\nu,eff}$ scales with 
the upper bound chosen for $\Omega_{\rm M}$ (\ie interesting CMB bounds on 
the RED require a prior constraint on $\Omega_{\rm M}$).  

\subsection{\label{sub:SNIa}Constraints from Type Ia Supernovae}

Given a distant population of so-called ``standard candles'', the 
magnitude-redshift relation is a powerful way to determine cosmological
parameters directly \cite{SAND61}.  In a standard FRW cosmology, the apparent 
bolometric magnitude, $m(z)$, of a standard candle is related to its absolute 
bolometric magnitude, $M$, and redshift by
\begin{widetext}
\beq
\label{eq:eq7}
m(z) = M+ 5\log({\cal D}_L(z, \Omega_{\rm M}, \Omega_{\rm \Lambda}, \Omega_{\rm R}))
-5\log(H_0)+25,
\eeq 
where $H_0$ is the Hubble parameter in kms$^{-1}$Mpc$^{-1}$.
${\cal D}_L$ is the ``Hubble-constant-free'' luminosity distance in Mpc, given by 

\begin{displaymath}
{\cal D}_L = c(1+z)\sqrt{\frac{1}{|\Omega_{\rm k}|}} \Sigma \Bigg( 
\sqrt{|\Omega_{\rm k}|} \int_{0}^{z}\frac{d \overline z}{\sqrt{(1+ \overline z)^2
(1+\Omega_{\rm M}\overline z)+(1+ \overline z)^2(2+\overline z) \overline z
\Omega_{\rm R} - \overline z(2+\overline z)\Omega_{\rm \Lambda}}} \Bigg)
\nonumber
\end{displaymath}
\end{widetext}
where $c$ is in kms$^{-1}$, $\Omega_{\rm k} \equiv 1-\Omega_{\rm M}-
\Omega_{\rm \Lambda}-\Omega_{\rm R}$, and

\begin{displaymath}
\begin{array}{ccl}
\Sigma(x) & = & \sin(x) \quad {\rm if} \quad \Omega_{\rm k} < 0 \nonumber \\
\quad     & = & x \quad {\rm if} \quad \Omega_{\rm k} = 0 \nonumber \\
\quad     & = & \sinh(x) \quad  {\rm if} \quad \Omega_{\rm k} > 0. 
\end{array}
\end{displaymath}
The matter, radiation and cosmological constant energy densities enter 
${\cal D}_L$ with different powers of $z$, making it possible to utilize 
observations of standard candles over a range of redshifts to determine 
the cosmological density parameters (for a review, see \cite{GOOB}).

Two groups, the Supernova Cosmology Project \cite{SCP} and 
the High-z Supernova Search Team \cite{HZT}, have been engaged 
in a systematic study of the magnitude-redshift relation of high-redshift, 
type Ia supernovae in an effort to constrain $\Omega_{\rm M}$ and 
$\Omega_{\rm \Lambda}$.  We have re-analyzed the data of 
Perlmutter et al. \cite{SCP} allowing for a non-negligible contribution from 
$\Omega_{\rm R}$.  For simplicity, we have assumed flatness to be a robust 
result of CMB measurements \cite{JAFFE} and have performed a 
maximum-likelihood analysis (we take ${\cal L} \propto e^{-\chi^2/2}$ subject to the 
priors $\Omega_{\rm M} \geq 0$, 
$\Omega_{\rm R} \geq 0$) in order to derive constraints on \Om and \Or with 
$\Omega_{\rm \Lambda} \equiv 1-\Omega_{\rm M}-\Omega_{\rm R}$\footnote{Relaxing 
flatness yields a less restrictive bound on $\Omega_{\rm R}$ but much of the 
favored region of parameter space would be ruled out by age considerations, CMB 
measurements, or the aforementioned bounds on $\Omega_{\rm M}$.}.  

\begin{figure}
\resizebox{!}{8cm}
{\includegraphics{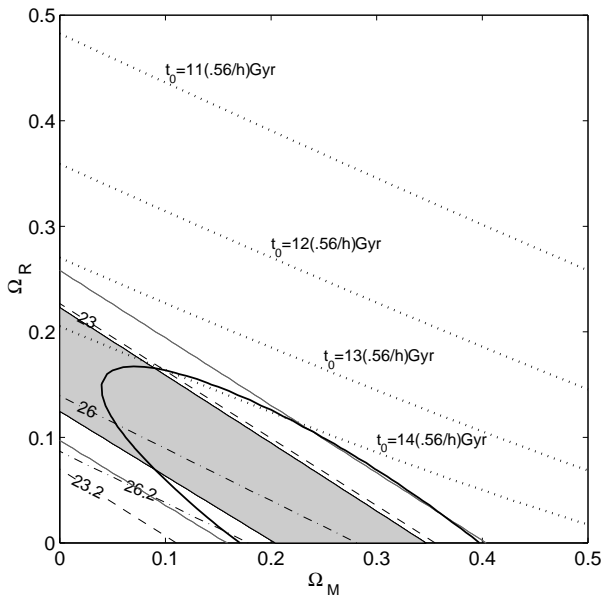}}
\caption{\label{fig3} 68\% (filled) and 95\% (open) confidence contours 
in the $\Omega_{\rm M}$-$\Omega_{\rm R}$ plane.  For illustration, we show several 
labeled isochrones (dotted) and the elliptical 95\% confidence contour obtained by 
adding an additional prior constraint of $\Omega_{\rm M} = 0.3 \pm 0.1$ (solid, heavy).  
Also shown are contours of constant effective ${\rm m}_B$ for SNIa at redshifts of $z=0.5$ 
(dashed at ${\rm m}_B = 23,\ 23.2$) and $z=1.5$ (dash-dot at ${\rm m}_B = 26,\ 26.2$).}
\end{figure}

In Figure \ref{fig3}, we show the constraints on $\Omega_{\rm M}$ and 
$\Omega_{\rm R}$ from SNIa.  The projection $\Omega_{\rm R}$ = 0 is consistent
with earlier analyses that found $\Omega_{\rm M} \sim 0.3$ and $\Omega_\Lambda 
\sim 0.7$ assuming the RED to be negligible.  Allowing for relativistic energy density, 
we find a degenerate set of $\Omega_{\rm M}$ and $\Omega_{\rm R}$ that are 
consistent with the SNIa data: the high-matter-content (\ie 30\% matter, 
70\% cosmological constant) flat Universe and the high-radiation-content (\ie 
20\% radiation, 80\% cosmological constant) flat Universe are equally good fits.  
Observe that regardless of whether relativistic energy is allowed or not, the SNIa
data require a large cosmological constant ($0.6\lsim \Omega_\Lambda\lsim 0.9$).

\begin{figure}
\resizebox{!}{7.5cm}
{\includegraphics{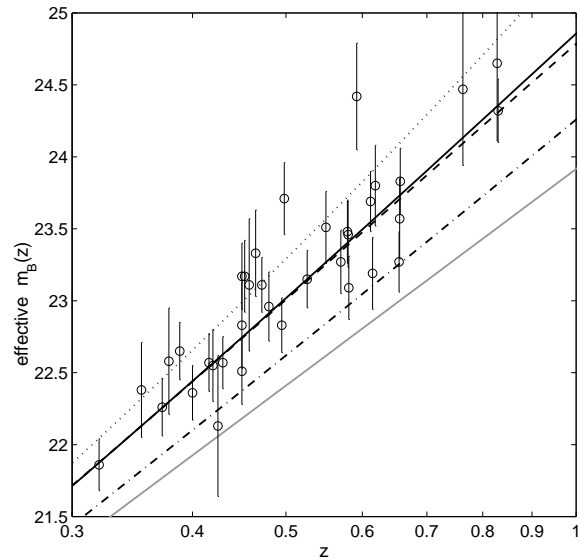}}
\caption{\label{fig4} The high-redshift portion of the Hubble diagram
for the SCP data (with $1\sigma$ error bars) alongside the 
magnitude-redshift relation in several flat
cosmologies:  $\Omega_{\rm M}=0.3$ and $\Omega_{\rm R}=0$ (heavy, solid),   
$\Omega_{\rm M} = 0$ and $\Omega_{\rm R} = 0.2$ (dashed), 
$\Omega_{\rm \Lambda} = 1$ (dotted), $\Omega_{\rm M} = 1$ (dash-dot), 
$\Omega_{\rm R} = 1$ (light, solid).}
\end{figure}

The 95\% confidence contour is approximately fit by the region\beq
\label{eq:eq8}
\Omega_{\rm R} + 0.62\Omega_{\rm M} \approx 0.17 \pm 0.08.
\eeq
It is not surprising that the data pick out this degenerate valley in the 
$\Omega_{\rm M}$-$\Omega_{\rm R}$ plane because most of the SNIa data are 
from $z\sim 0.5$ and our degenerate valley represents the parameters with 
approximately constant luminosity distance at this redshift.
The degeneracy can be further understood with the help of 
Figures \ref{fig3} \& \ref{fig4}.  Dashed and dot-dashed lines in 
Figure \ref{fig3} depict contours of constant apparent magnitude at 
redshifts $z=0.5$ and $z=1.5$ respectively.  As most of 
the SNIa data lie near $z \sim 0.5$, the confidence region is nearly 
parallel to lines of constant m$_B$ at $z=0.5$.  At higher redshifts, lines of 
constant apparent magnitude have a more shallow slope and are closer together, 
thus observations of SNIa at $z>0.5$ can break the matter-radiation degeneracy.   

Figure \ref{fig4} further illustrates the large lever arm of high-$z$ SNIa for 
cosmological parameter estimation.  Notice that the magnitude-redshift relation in a 
high-matter-content Universe ($\Omega_{\rm M}=0.3$, $\Omega_{\rm R}=0$) 
is quite similar to that in a high-radiation-content Universe 
($\Omega_{\rm R}=0.2$, $\Omega_{\rm M}=0$) at $z\lsim0.5$.  At higher redshift,
one begins to probe the epoch prior to matter-$\Lambda$ and/or radiation-$\Lambda$
equality; the cosmological constant becomes increasingly unimportant
compared to radiation and/or matter and the two curves begin to diverge.  
The SCP data only extend to $z=0.83$, therefore they cannot be used to
scrutinize this earlier phase and they cannot distinguish between a 
high-matter-content Universe and a high-radiation-content Universe.  
The proposed Supernova Acceleration Probe ([http://snap.lbl.gov]) may 
have the ability to break this degeneracy by observing many more SNIa at 
significantly higher redshift.

Marginalizing over $\Omega_{\rm M}$ by integrating the likelihood, we obtain
a bound on the RED at low redshift $\theta_{\rm SNIa} \equiv \theta(z\lsim0.5)$,

\beq
\label{eq:eq9}
\theta_{\rm SNIa} \lsim 3.4h^2\times 10^3\ (68\%)\textrm{ or }4.8h^2\times 10^3\ (95\%).  
\eeq
In terms of the RED today, the bound is $\Omega_{\rm R} \le 0.14$ (68\%) and 
$\Omega_{\rm R} \le 0.20$ (95\%).  

The SNIa constraint on the RED applies during recent epochs, namely, $z \lsim 0.5$, 
as can be seen in Figure \ref{fig2}.  Of course, in light of other estimates of
cosmological parameters, the allowed region in Fig. \ref{fig3} is not equally probable.
Estimates of the contemporary matter density of the Universe favor the 
region $\Omega_{\rm M} = 0.3 \pm 0.1$ and would slightly reduce the 
SNIa upper bound on the RED.  This is illustrated in Figure \ref{fig3} where
we also plot  the 95\% confidence contour obtained with the additional prior
constraint $\Omega_{\rm M}=0.3\pm0.1$.  This prior constraint on 
the matter density reduces the 95\% bound on $\Omega_{\rm R}$ by about 25\%.  

Note also that the SNIa bound on the contemporary RED
is more stringent than bounds that follow from the requirement that the 
expansion age of the Universe be at least as large as the ages of the oldest objects in the
Universe.  In Figure \ref{fig3}, the {\it entire} region delineated by SNIa 
data corresponds to an acceptable age.  In particular, a high-radiation-content 
Universe with $\Omega_{\rm M} = 0.04$, $\Omega_{\rm R} = 0.20$, 
$\Omega_{\rm \Lambda} = 0.76$ and a Hubble parameter at the extreme lower limit, 
$h=0.56$, is $13.8\textrm{ Gyr}$ old.  Even with $h=0.72$, $t_0 
\simeq 10.7\textrm{ Gyr}$, a value that is not grossly inconsistent with the ages of the 
oldest stars and globular clusters \cite{AGE}.

\subsection{\label{sub:LSS}Large Scale Structure Constraints}

The effect of additional relativistic energy on the growth of large scale 
structure (LSS) has been studied by numerous authors \cite{TSK84, ST85, LSSSIM, MPBS, TW97}.  
Typically, the introduction of ``hot dark matter''
was considered a means to suppress power on small scales and thus 
reconcile an $\Omega_{\rm M} = 1$, Cold Dark 
Matter (CDM) cosmology with the observed power spectrum derived from galaxy 
surveys \cite{SPECT}.  Conversely, too much relativistic energy
adversely affects the growth of structure and therefore LSS can be
used to constrain the RED.  In light of mounting 
evidence, the CDM paradigm has given way to the so-called \lcdm paradigm 
with $\Omega_{\rm M} \sim 0.3$ and $\Omega_{\rm {\Lambda}} \sim 0.7$.  With this
in mind, we revise previous work in order to constrain the relativistic 
energy content of the Universe using LSS.

The most striking feature of a CDM or \lcdm type power spectrum is a break 
in the power law at, roughly speaking, the co-moving horizon scale at 
matter-radiation equality

\beq
\label{eq:eq10}
\lambda_{\rm EQ} \simeq 16(\Omega_{\rm M}h)^{-1}\ h^{-1}\textrm{ Mpc}.
\eeq
This feature arises because the growth of sub-horizon sized perturbations
is quelled by the cosmological expansion during radiation domination.  Hence, 
perturbations on scales smaller than $\lambda_{\rm EQ}$ are suppressed by 
a factor $\sim (\lambda/\lambda_{\rm EQ})^2$ relative to scales that were 
super-horizon sized at matter radiation equality.

If the RED of the Universe is contained entirely
in photons and three light neutrino species, and if the primordial
power spectrum is nearly scale-invariant, the \lcdm matter power spectrum 
can be expressed in terms of only one quantity, the shape parameter $\Gamma \simeq
\Omega_{\rm M}h$ \cite{BBKS} (this neglects the 
effect of baryons, see \cite{GBAR}).  
Several authors have used LSS observations on linear scales to infer 
acceptable values of the shape parameter \cite{SHAPE}.  
One of the more permissive of these determinations is the $95\%$ range 

\beq
\label{eq:eq11}
0.06 \le \Gamma \le 0.46,
\eeq
quoted by Efstathiou \& Moody \cite{EM}, which we will use to constrain 
the RED.  

In the presence of excess relativistic energy, the horizon scale at 
matter-radiation equality is no longer given by (\ref{eq:eq10}).  Rather, 

\beq
\label{eq:eq12}
\lambda_{\rm EQ} \simeq 16(\Omega_{\rm M}h)^{-1}
\sqrt{\Omega_{\rm R}/\Omega_{\rm \gamma\nu}}h^{-1}\textrm{ Mpc}
\eeq
and the effective shape parameter is therefore given by \cite{LSSSIM}

\beq
\label{eq:eq13}
\Gamma \simeq \Omega_{\rm M}h\theta^{-1/2}.
\eeq
Taking the lower bound $\Gamma > 0.06$  
we immediately come upon a generic constraint on the mean RED:

\beq
\label{eq:eq14}
\theta \le \Bigg(\frac{\Omega_{\rm M}h}{0.06}\Bigg)^2.
\eeq
With our conservative assumptions that $\Omega_{\rm M} \le 0.5$ and $h \le 0.88$,
the corresponding restriction on the RED during (and prior to) the epoch of
matter-radiation equality, $\theta_{\rm LSS} \equiv \theta(z=z_{\rm EQ})$, is

\beq
\label{eq:eq15}
\theta_{\rm LSS} \lsim 54
\eeq
where $1+z_{\rm EQ} \equiv \Omega_{\rm M}/\Omega_{\rm R}$.  The 
relative weakness of this bound is due to our conservative 
lower bound on $\Gamma$.  Taking the 95\% band of Eisenstein \& 
Zaldarriagga \cite{SHAPE}, $0.15 \le \Gamma \le 0.58$, results in 
$\theta_{\rm LSS} \lsim 9$.

Note that this constraint (\ref{eq:eq15}) allows the epoch of matter-radiation equality 
to be at redshifts as low as $z_{\rm EQ} \approx 120$.  We can obtain 
a more stringent bound on the RED by following an argument invoked by Turner, 
Steigman \& Krauss \cite{TSK84}, Steigman \& Turner \cite{ST85} and 
Turner \& White \cite{TW97}.  Assuming 
a nearly scale-invariant primordial power spectrum, data from 
the COBE DMR experiment \cite{COBEDMR} indicate that 
the rms density contrast at horizon crossing is on the order 
of $\delta_{\rm H} \sim \textrm{few} \times 10^{-5}$ \cite{DELH}.  
Meanwhile, measurements of the galaxy correlation function reveal 
nonlinear clustering on scales smaller than a critical scale, 
$\lambda_{\rm NL} \sim 5 h^{-1}\textrm{ Mpc}$ \cite{XI_R}.  
We adopt the conservative constraint that perturbations on 
scales smaller than $\lambda_{\rm NL}$ must have grown by at least a factor of 
$\gamma_{\rm min} \equiv 10^{3}$ in order for the rms perturbation on these 
scales to be nonlinear.  In linear perturbation 
theory, density fluctuations grow as $\delta \propto (1+z)^{-1}$ 
during matter domination and only logarithmically during radiation domination.  
This implies that the matter dominated epoch must span at least three
orders of magnitude in redshift.  With $\Omega_{\rm M} \le 0.5$ this, in
turn, imposes the limitation 

\beq
\label{eq:eq16}
\theta_{\rm LSS}h^2 \lsim 12 \textrm{   or   } \theta_{\rm LSS} \lsim 38,
\eeq
where we have taken $h \ge 0.56$.  This constraint is shown in Figure 
\ref{fig2} where we summarize the constraints on relativistic energy
imposed by BBN, the CMB, LSS and the SNIa magnitude-redshift relation.  

\section{\label{sec:WIMP}Constraining Relic Decays}

In the preceding section we reported limits on the RED at various epochs.  
In the absence of electron neutrino degeneracy, the BBN constraint on the RED is, 
by far, the most stringent; it leaves little room for any non-standard 
relativistic energy during the BBN epoch.  One way to circumvent the BBN constraint 
is to inject relativistic energy after nucleosynthesis has ended, for instance, 
by the decay of a particle that was non-relativistic during BBN 
($M \gg 1\textrm{ MeV}$) into relativistic products.  As such, the study of relic 
particle decays comes part and parcel with constraints on relativistic energy.  
We examine the simple case of a massive, unstable big bang relic which may be 
pertinent to physics ``beyond the Standard Model'' and discuss constraints on the 
relic's mass and lifetime that follow from the bounds on the RED in Section 
\ref{sec:CRE}.  It is important to note that, 
aside from the CMB constraint, the assumption that the relic is very massive is 
not critical.  In general, the particle must be non-relativistic at decay in order 
for its decay product's energy density to be comparable to or greater than 
$\rho_{\rm \gamma\nu}$.

Consider the decay of a massive relic particle X, with lifetime $\tau$, 
into relativistic products.  Had X not decayed, the energy density in 
these particles today would be

\beq
\label{eq:eq17}
\Omega_{\rm X}h^{2} \simeq 274M_{keV}Y.
\eeq
In (\ref{eq:eq17}), $M_{keV}$ is the mass of the particle in keV and $Y$ is the ratio
of the number density of the particle to the entropy density ($Y \simeq 0.039$
for a light neutrino, $Y \simeq 2 \times 10^{-10}$ for a 5 GeV, Dirac neutrino).  Given 
a specific particle physics model in which X is produced in the early Universe, 
$Y$ is fixed (in the absence of subsequent entropy production); however, in order to 
make the constraints on heavy particle decays as generic as possible, we have chosen 
to keep $Y$ as an explicit, free parameter.  
Further, we assume that the daughter particles are weakly-interacting 
(constraints on radiative decays are quite severe \cite{RAFFELT}).  In all 
analytic calculations we assume that decays occur simultaneously at $t=\tau$.  
Energy conservation during decay demands that the present energy density in decay 
products be $\Omega_{\rm D}= \Omega_{\rm X}/(1+z_{\rm D})$, where $z_{\rm D}$ is 
the redshift at decay (\ie $z$ at $t=\tau$).  Assuming the Universe to be X-dominated 
prior to decay, 

\beq
\label{eq:eq18}
\Omega_{\rm D}h^{2} \simeq 5.1 \times 10^{-4}M_{keV}^{4/3}Y^{4/3}
\tau_{yr}^{2/3}
\eeq
where $\tau_{yr}$ is the lifetime of X in years.  

\subsection{\label{sub:CMBWIMP}CMB Constraints on Relic Properties}

\begin{figure}
\resizebox{!}{6.5cm}
{\includegraphics{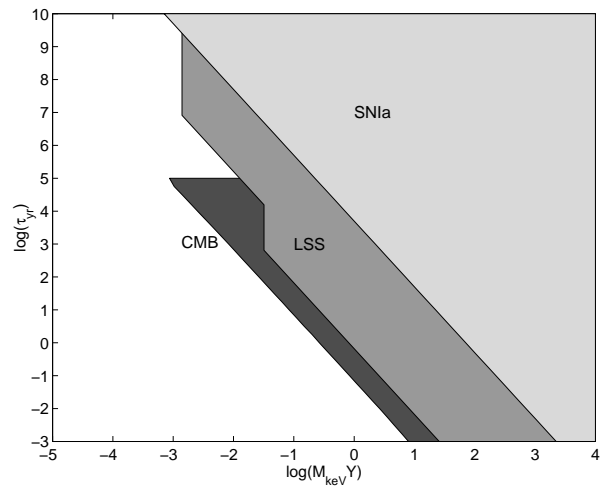}}
\caption{\label{fig5} The regions of parameter space for a massive relic decay that 
are excluded by CMB (heavily shaded), LSS (moderately shaded) and SNIa (lightly shaded) 
arguments.  Each excluded region extends to arbitrarily high $M_{keV}Y$.}
\end{figure}

One constraint on relic properties follows from the requirement that 
the total energy density in relativistic particles $\Omega_{\rm R}
= \Omega_{\rm \gamma\nu} + \Omega_{\rm D}$, fall under the CMB bound in 
(\ref{eq:eq6}) during the epoch of recombination.  If the particle 
decays prior to recombination the CMB constraint (\ref{eq:eq6}), 
along with the analytic approximation of (\ref{eq:eq18}), implies 
that\footnote{The CMB constraint on relativistic energy is strict enough 
that the assumption of X-domination prior to decay is untenable, making it 
necessary to integrate the equations governing heavy WIMP decay in a 
flat, Friedmann cosmology.  We have performed the necessary integration 
and find that the above analytic bound is typically accurate to 
within 20\%.}   

\beq
\label{eq:eq19} 
M_{keV}^2Y^2\tau_{yr} \lsim 7.5 \times 10^{-2}. 
\eeq
For pertinent lifetimes, $10^{-3} \lsim \tau_{yr} \lsim 10^5$, the excluded 
region of parameter space is displayed in Figure \ref{fig5}.

Post-recombination X decays modify the CMB power spectrum through the ISW effect 
and through shifts in the multipole positions of the acoustic peaks due to a change in the 
angular diameter distance to the surface of last-scattering.  It is possible to 
use this modification of the observed CMB anisotropy power spectrum to constrain 
post-recombination X decays \cite{LDCMB}, but these constraints 
would not be generic because the gravitational dynamics of the relic as well as its decay
scheme can contribute to the ISW effect.  Such constraints would have to be developed on a 
case-by-case basis considering the relic mass and relic abundance separately; however, we 
mention a specific case that evades the CMB bound and contributes a significant RED at the 
present epoch in section \ref{Con}.  In the following subsection, we show that the growth of 
LSS can place severe, yet generic, constraints on post-recombination decays.  

\subsection{\label{sub:LSSWIMP}LSS Constraints on Relic Decays}

Large scale structure considerations also lead to bounds on decaying relics.
Utilizing (\ref{eq:eq18}), $\theta$ can be expressed as 
$\theta \simeq 1 + 12.2M_{keV}^{4/3}Y^{4/3}\tau_{yr}^{2/3}$.
Consequently, the effective shape parameter (\ref{eq:eq13}) may be written

\beq
\label{eq:eq20}
\Gamma \simeq \Omega_{\rm M}h[1 + 12.2M^{4/3}_{keV}Y^{4/3}\tau_{yr}^{2/3}]^{-1/2}
\eeq
from which we conclude that 

\beq
\label{eq:eq21}
M^{2}_{keV}Y^2\tau_{yr} \simeq 188[(\Omega_{\rm M}h)^2-\Gamma^2]^{3/2}.
\eeq
Again, taking $\Omega_{\rm M} \le 0.5$, $h \le 0.88$ and $\Gamma \ge 0.06$, 
we deduce that 

\beq
\label{eq:eq22}
M^2_{keV}Y^2\tau_{yr} \lsim 16.
\eeq

Observe that (\ref{eq:eq22}) restricts the combination 
$M^{2}_{keV}Y^{2}\tau_{yr}$.  As has been underscored by McNally \& Peacock 
and Bharadwaj \& Sethi \cite{MPBS}, the power spectrum will also exhibit a feature 
on small scales corresponding to the transition between radiation domination and 
an early matter dominated (domination by X particles) phase.  As the co-moving 
horizon scale during the first epoch of matter-radiation equality is given by

\beq
\label{eq:eq23}
\lambda_{\rm X} \simeq \frac{1}{17.1M_{keV}Y}\textrm{ Mpc},
\eeq
this small-scale feature can, in principle, constrain the combination 
$M_{keV}Y$ alone.  In practice, however, much of the interesting region of
relic parameter space would correspond to nonlinear scales and such a constraint 
would require a better theoretical handle on nonlinear clustering and bias.

We can strengthen the bound in (\ref{eq:eq22}) by requiring that structure 
grow sufficiently.  As Steigman and Turner \cite{ST85} have noted, there 
are two ways in which this can occur.  One way was mentioned in Section 
\ref{sub:LSS}, namely, that the relic decay early enough so that the most recent 
epoch of matter domination began at a redshift $(1+z_{\rm EQ}) \ge 10^{3}$.  The 
redshift of equality is $(1+z_{\rm EQ}) \simeq \Omega_{\rm M}/\Omega_{\rm D}$ and 
using (\ref{eq:eq17}) and (\ref{eq:eq18}) we find that this scenario requires

\beq
\label{eq:eq24}
M^2_{keV}Y^2\tau_{yr} \lsim 0.66
\eeq
and is relevant for lifetimes in the range $10^{-3} \lsim \tau_{yr} \lsim 10^{5}$.

In the presence of a massive unstable relic however, there can be 
two phases of matter domination, an early X
dominated phase and a second matter dominated phase after relic decay.  
It may be possible for perturbations on scales $\lambda < \lambda_{\rm NL}$ 
to take advantage of both of these periods of growth and thereby grow by a 
factor greater than $\gamma_{\rm min}$.  X domination begins at redshift 
$(1+z_{\rm X}) = \Omega_{\rm X}/\Omega_{\rm \gamma\nu}$, followed by 
decay at redshift $(1+z_{\rm D}) = \Omega_{\rm X}/\Omega_{\rm D}$.  
Thus the total growth factor for scales {\it smaller} than the horizon scale
at X domination, $\lambda_{\rm X}$, is

\beq
\label{eq:eq25}
\gamma \simeq \frac{(1+z_{\rm X})}{(1+z_{\rm D})}(1+z_{\rm EQ}) \simeq
2.4 \times 10^4\Omega_{\rm M}h^2.
\eeq 
With our aforementioned limits on \Om and $h$, these scales grow by a maximum 
of $\gamma \simeq 9.3 \times 10^3$.  Perturbations that enter the
horizon after X domination begins, grow by a smaller factor: 

\beq
\label{eq:eq26}
\gamma(\lambda>\lambda_{\rm X}) \simeq 9.3 \times 10^3 \Bigg(\frac{\lambda_{\rm X}}{\lambda}\Bigg)^2.
\eeq
Compelling $\gamma(\lambda_{\rm NL})$ to be greater than $\gamma_{\rm min}$ forces

\beq
\label{eq:eq27}
M_{keV}Y \lsim 3.6h \times 10^{-2} \lsim 3.2 \times 10^{-2}.
\eeq
Combining (\ref{eq:eq22}), (\ref{eq:eq24}) and (\ref{eq:eq27}), we summarize
the LSS constraints on early relic decays as 

\begin{eqnarray}
\label{eq:eq28}
M^2_{keV}Y^2\tau_{yr} \lsim 0.66 \quad \textrm{or} & \nonumber \\
M^2_{keV}Y^2\tau_{yr} \lsim 16 \quad \textrm{and} & M_{keV}Y \lsim 3.2 \times 10^{-2}.
\end{eqnarray}
The excluded region is shown in Figure \ref{fig5}.  

The above constraints (\ref{eq:eq28}) are pertinent for lifetimes in the 
range $10^{-3} \lsim \tau_{yr} \lsim 10^6$ because we assumed that 
the turnover in the power spectrum is indicative of $\lambda_{\rm EQ}$, the horizon 
scale at the epoch of matter-radiation equality {\it after} X decay.  
Alternatively, the X particles may have very large lifetimes, $\tau_{yr} \gg 10^{6}$, 
in which case the turnover in the power spectrum would be indicative of 
$\lambda_{\rm X}$, the horizon scale at the first epoch of matter-radiation equality, 
{\it prior to} X decay.  In this case, the effective shape parameter is simply

\beq
\label{eq:eq29}
\Gamma \simeq (\Omega_{\rm X} + \Omega_{\rm M})h.
\eeq
If we adopt our limiting case that $\Omega_{\rm M} \ge 0.1$ in order to 
be consistent with various measures of the contemporary matter density, we find 
that the restriction $\Gamma \le 0.46$ from (\ref{eq:eq11}) asserts that

\beq
\label{eq:eq30}
M_{keV}Y \lsim 1.2 \times 10^{-3}.
\eeq
We illustrate this bound in Figure \ref{fig5} by the vertical boundary
for $\tau_{yr} \gsim 10^{7}$.  Lastly, because $(1+z_{\rm EQ}) \sim 10^4$
in this scenario and $\Omega_{\rm X}$ is not more than a factor of five
larger than our lower bound on \Om, requiring $\gamma_{min} \gsim 10^3$ 
provides only a weak restriction on X lifetimes.  It may be more useful to 
take advantage of the large scale feature that would be present in 
the power spectrum due to the injection of relativistic energy in order to 
limit relic properties.  We do not explore such bounds as this would require 
specifying $M_{keV}$ and $Y$ separately.  

\subsection{\label{sub:SNWIMP}SNIa Constraint on Relic Decays}

With the SNIa bound on relativistic energy from Section \ref{sub:SNIa}, it is 
now easy to obtain a SNIa bound on the relic decay properties.  Using (\ref{eq:eq18}) 
and the 95\% upper limit in (\ref{eq:eq9}), we have

\beq
\label{eq:eq31}
M_{keV}^2Y^2\tau_{yr} \lsim 5.3 \times 10^3,
\eeq
where we have once again assumed $h\le0.88$.  Notice that this follows from a 
bound on the contemporary RED and, therefore, is most pertinent to late  
decays (\ie $\tau_{yr} \gsim 10^8$).  Again, this constraint is shown in Figure 
\ref{fig5} where our bounds on the properties of decaying relics are summarized.  

\section{\label{Con}Conclusions}

Constraints on the cosmological RED can provide a fundamental probe of 
particle physics beyond the standard model.  In this paper we have discussed
constraints on the RED during four distinct epochs arising from BBN, the CMB, 
LSS and SNIa (see Figure 2).  Further, we have shown how these bounds constrain the mass and
lifetime of a hypothetical big bang relic (see Figure 5).  Somewhat surprisingly, the RED at 
the current epoch is relatively unconstrained: we have shown that the 
magnitude-redshift relation for SNIa is consistent with a flat universe comprised 
of up to 20\% relativistic energy.  Conventional wisdom suggests that the RED 
today must be small to allow for sufficient growth of large scale structure and not appreciably 
alter the CMB anisotropy power spectrum.  The LSS bound does, 
in fact, significantly limit the RED {\it near the epoch of matter-radiation equality.}  
Any RED consistent with subsequent growth of LSS, redshifted to the current epoch, 
would be quite small.  Conversely, a RED that is large yet acceptable with respect 
to the SNIa bound would clearly inhibit the growth of LSS if it were redshifted to 
the past.  However, a long-lived particle that decays sufficiently late as to
avoid the LSS constraint could nevertheless contribute substantially to the RED today.
In this case, the relevant constraint would be the aforementioned CMB bound (see Section \ref{sub:CMBWIMP}).  
Avoiding the CMB constraint requires the X particles to be very long-lived if they are to 
contribute appreciably to the RED today.  
Consider a big bang relic with $M_{keV}Y = 1.2 \times 10^{-3}$(a 30 eV neutrino is an example of a
particle with the necessary abundance) and a very long lifetime, $\tau_{yr} = 2 \times 
10^9$.  The decay products of this relic contribute $\Omega_{\rm D}h^{2} \sim 0.1$ 
and its properties are marginally consistent with the growth of LSS .  
In addition, the decay products are produced sufficiently late so as to contribute an 
unconstrained ISW perturbation at low multipole moments, peaking around 
$\ell \sim 10$, and to change the angular diameter distance to the last-scattering 
surface by only $\sim 7\%$ (see Figures 3 and 4 of Kaplinghat \etal \cite{LDCMB}).  These effects 
cannot be ruled out by current CMB data.  In Figure \ref{fig2} we show the evolution of the 
RED including the decay products of this hypothetical, long-lived big bang relic.

\begin{acknowledgments}

We would like to thank Jim Kneller, Savvas Koushiappas, Bob Scherrer, and Gary 
Steigman for many insightful comments and helpful discussions. We also wish to thank 
Rich Schugart and Jason Farris who performed some initial SNIa calculations as part of 
the NSF REU program at OSU.  We acknowledge DOE contract DE-FG02-91ER40690 for support.

\end{acknowledgments}

\end{document}